\documentclass[aps,twocolumn,epsfig,eqsecnum]{revtex4}
\usepackage{graphicx}
\newcommand\be{\begin{eqnarray}}
\newcommand\ee{\end{eqnarray}}
\newcommand\ba{\begin{array}}
\newcommand\ea{\end{array}}

\def\r{\rangle}
\def\l{\langle}
\def\T{{\rm Tr}}

\def\cH{{\cal H}}
\def\cS{{\cal S}}
\def\cP{{\cal P}}

\def\cI{{\cal I}}

\def\cF{{\cal F}}

\def\cE{{\cal E}}

\def\cA{{\cal A}}

\begin{document}

\title{Entanglement measures: State ordering {\it vs.} local operations}
\author{M\'ario Ziman and Vladim\'\i r Bu\v zek}
\affiliation
{Research Center for Quantum Information, Slovak Academy of Sciences, D\'ubravsk\'a cesta 9, 845 11 Bratislava, Slovakia \\
Faculty of Informatics, Masaryk University, Botanick\'a 68a, 602 00 Brno, Czech Republic\\
{\em Quniverse}, L{\'\i}\v{s}\v{c}ie \'{u}dolie 116, 841 04 Bratislava, Slovakia}

\begin{abstract}
A set of all states of a bi-partite quantum system can be divided into subsets each of which contains
states with the same degree of entanglement. In this paper
we address a question whether local operations
(without classical communication) affect the entanglement-induced
state ordering. We show that arbitrary unilocal channel (i.e., a channel that acts on one sub-system of a bi-partite system only)
might change the ordering for an arbitrary nontrivial measure of entanglement.
A slightly weaker result holds for the maximally entangled states. In particular, the maximally entangled states might not remain the most
entangled ones at the output of a unilocal noise channel.

\end{abstract}

\maketitle


\section{Quantum entanglement}
Quantum phenomena (such as quantum dense coding \cite{densecoding},
quantum teleportation \cite{teleportation}, quantum secret
sharing \cite{secretsharing}, etc.) associated with the existence of quantum entanglement
represent one of the most important pillars of quantum information theory \cite{nielsen}.
In spite of all the progress in understanding the nature of this phenomenon
some features of the concept of quantum entanglement are still to be properly illuminated.
In particular, due to the seminal work of Reinhard Werner
\cite{werner_ent} and others (see e.g. the review article \cite{review_ent}) we have a precise mathematical
definition of what does it mean when we say that a bi-partite state is entangled. On the other hand
a clear generally applicable operational meaning of the entanglement is still missing.

In this paper we will analyze some dynamical aspects of quantum entanglement.
Specifically we will study the relation between unilocal operations and static
(kinematic) properties of quantum entanglement expressed in terms of the entanglement-induced
state ordering.

The concept of quantum entanglement is relatively easy
to understand when we deal with pure states of bi-partite systems. This easiness originates in a close (mathematical)
relationship between the concept of entanglement and the concept of statistical correlations.
In fact, for pure quantum states these two concepts can be quantified by the same
functions and the meaning of the statement ``not entangled'' is equivalent to the ``not correlated''.
However, conceptual differences between entanglement and statistical correlations become striking
when we consider mixed states.

An important feature
of quantum entanglement reflecting its behavior under local operations
and classical communication has been known for some time \cite{nielsen}.
Namely, it is well established that two (classically) communicating
distant parties cannot entangle their quantum systems without
performing a global operation (corresponding to some effective interaction).
In other words, arbitrary local operations cannot create
the entanglement even if these actions are coordinated by
an exchange of classical information. Moreover,  local unitary
transformations do not affect the quantum entanglement at all.

These properties form a basis of our intuitive picture of quantum entanglement.
Let us summarize these ``natural'' properties of entanglement:
\begin{itemize}

\item The quantum entanglement is a property of a quantum state.

\item A quantum state is  entangled, if it cannot be prepared
from a factorized state ($\varrho_A\otimes\varrho_B$) by an action of local
operations and classical communication, i.e. it cannot be expressed
as a convex sum of factorized states ($\varrho_{AB}\ne \sum_j p_j \varrho_A^{(j)}\otimes\varrho_B^{(j)}$).

\item LOCC (local operations plus classical communication) operations
applied to an arbitrary (even entangled) quantum state can only destroy
the entanglement.

\item Locally unitary equivalent quantum states are equally entangled.
\end{itemize}

As we have already said, the concept of ``not being entangled'' is well defined. Non-entangled
states are called {\em separable}. There is also a common agreement on the notion
of maximally entangled quantum states that represent the other extreme.
We say that a bi-partite quantum state is maximally entangled if it is pure and the two
subsystems are in maximally mixed states, i.e.
$\varrho_{AB}=|\Psi\r\l\Psi|$ and
${\rm Tr}_B[|\Psi\r\l \Psi|]={\rm Tr}_A[|\Psi\r\l \Psi|]=\frac{1}{d}I$
with $d=\min{\{{\rm dim}\cH_A, {\rm dim}\cH_B\}}$.

There are two basic
questions: i) whether a given state is entangled, or not?, and
ii) whether we can compare the entanglement of different
quantum states. Both questions can be addressed via the so-called
{\em entanglement measures}.

In this paper we will focus our attention on the concept of entanglement measures. We will
study dynamics of entanglement under the action of local channels. Our paper is organized as follows:
We start with a brief introduction to entanglement measures. Then we will
analyze the stability of entanglement-induced state ordering and the
properties of maximally entangled states with respect to
local operations, in particular for the so-called unilocal channels.
Finally, we will discuss some conceptual consequences
of our analysis.
\section{Entanglement measures}
The entanglement (see a recent review \cite{plenio_virmani})
has been identified as the key ingredient in applications such
as the quantum teleportation, the quantum secret sharing, etc. However, it is also
known that the presence of entanglement itself does not guarantee the success of a protocol. For instance,
an arbitrary entangled state cannot be used for the teleportation. Even if
a state can be exploited for this protocol the success/rate of the teleportation depends on
the particular state. Hence, it seems there are states with different
``quality'' and ``quantity'' of entanglement. In order to quantify a degree of entanglement
entanglement measures have been introduced. These measures
are functionals defined on a state space designed to quantify the
amount of entanglement in a given state. During the last ten years the
topic of entanglement measures has attracted a lot of attention and many
important results has been discovered.

Principally there are two approaches to the entanglement measures:
i) the {\em operational} approach, and ii)  the {\em axiomatic} approach.
The aim of the first approach is to
adopt a procedure (protocol) that crucially depends on the presence of entanglement (for example the quantum teleportation),
and to quantify its success of performance
depending on the particular state. Such measure would give
a direct operational meaning to quantum entanglement associated with a given state. Unfortunately no such (universal) procedure
is known. In the abstract axiomatic approach we reformulate our
intuitive understanding of entanglement into several axioms.
There exist several different (not completely equivalent) choices
for the system of axioms \cite{horodecki}, however our aim is not
to discuss all these choices. We say that the functional
$E:\cS(\cH)\to [0,\infty]$
is an entanglement measure if the following properties hold:
\begin{enumerate}
\item {\it Sharpness:}
$E(\varrho_{AB})=0$ if and only if $\varrho_{AB}$ is a separable state.
\item {\it Local unitary invariance:}
$E(U_A\otimes U_B\varrho_{AB}U_A^\dagger\otimes U_B^\dagger)=E(\varrho_{AB})$ for all unitary transformations $U_A,U_B$ and all states $\varrho_{AB}$.
\item {\it Normalization:}
$E(\varrho_{AB})$ is maximal only for maximally entangled states, i.e.
$E(\varrho_{AB})=\max_{\varrho_{AB}}E(\varrho_{AB})$ if and only if
$\T_A\varrho_{AB}=\T_B\varrho_{AB}\sim I$ and $\T\varrho_{AB}^2=1$.
\item {\it Nonincreasing under LOCC:}
A general LOCC operation transforms the original state $\varrho_{AB}$
into a mixture of states
$\omega_k^{AB}=\cE_k^{A}\otimes\cE_k^{B}[\varrho_{AB}]$
with probabilities $p_k$. This condition guarantees that the entanglement
is (on average) not created by LOCC operations, i.e.
$E(\varrho_{AB})\ge \sum_k p_k E(\omega_k^{AB})$.
\item {\it Convexity:}
$E(\sum_k p_k \omega_k^{AB})\le \sum_k p_k E(\omega_k^{AB})$.
\item {\it Additivity on pure states:}
$E(\Psi_{AB}\otimes\Phi_{A^\prime B^\prime})
=E(\Psi_{AB})+E(\Phi_{A^\prime B^\prime})$
for all pure states $\Psi_{AB},\Phi_{A^\prime B^\prime}$.
\end{enumerate}

The first four properties from the above list are in an agreement with our intuitive picture
discussed in the previous section. In order to motivate the remaining two
properties we have to take into account a situation in which
a pair of systems is a part of a larger composite object.
Without the loss of generality we can assume
to have three parties (systems) $A,B,C$ in a pure state $\Omega_{ABC}$. By performing
measurement on the system $C$ and reading an outcome $j$ (associated
with the state transformation $\cI_{AB}\otimes\cF_j^{C}$)
the original state $\varrho_{AB}=\T_{C}\Omega_{ABC}$ is transformed
into the state $\omega_j^{AB}=\T_C \Omega_j^{ABC}
=\T_C (\cI_{AB}\otimes\cF_j^C)[\Omega_{ABC}]$. This happens
with some probability $p_j$. Without the knowledge of the
observed outcome $j$, the experimentalists possessing the systems $A$ and $B$
can use only the entanglement contained in the state $\varrho_{AB}$, because
the measurement performed on $C$ does not affect the {\em average} state $\varrho_{AB}$.
However, if they acquire the information about the outcome $j$, they can exploit
the entanglement shared in particular states $\omega_j^{AB}$, hence they can
on average exploit $\sum_j p_j E(\omega_j^{AB})$ of the entanglement.
The knowledge of $j$ cannot decrease the entanglement contained originally
in $\varrho_{AB}$. Hence, although the measurement on the system $C$ is
a local action, the entanglement between $A$ and $B$ can increase,
i.e. the third party can assist to $A$ and $B$ to increase the
entanglement they share providing that the information on $j$
is communicated to $A$ and $B$. In fact,
the measurements on the system $C$ induces convex decompositions of the
state $\varrho_{AB}=\sum_j p_j\omega_j^{AB}$, thus we get the convexity
condition for entanglement measures.

For example, let us consider three parties $A,B$ and $C$ share a GHZ state
$|\Omega_{ABC}\r=\frac{1}{\sqrt{2}}(|000\r+|111\r)$. A bi-partite density operator
$\varrho_{AB}$ describes a classically maximally correlated state, which is not
entangled at all and cannot be used for the teleportation. On the other hand, if the third party $C$ performs  a measurement in
the dual basis $|\pm\r_C=\frac{1}{\sqrt{2}}|0\r\pm |1\r$ then for both outcomes $\pm 1$
the parties $A$ and $B$ share a maximally entangled quantum state.
In particular, $\omega_{\pm}^{AB}=\frac{1}{2}(|00\r\pm|11\r)(\l 00|\pm\l 11|)$.
We see explicitly that such an assistance by the third party
can significantly increase the entanglement - this is the reason for the
convexity condition. Taking the maximum of average entanglement
over all decompositions we obtain the so-called {\em entanglement of assistance} \cite{divincenzo},
$E_{\rm assist}(\varrho_{AB})=\max\sum_j p_j E(\omega_j^{AB})$.

The requirement of the additivity is a rather natural property of the quantum
entanglement, however we lack some clear operational reason for it
and it is not trivially satisfied for the measures we use.
For example, the additivity of entanglement
of formation is one of the most important open problems
in the quantum information theory. Therefore, it is demanded that
this property holds only for tensor product of pure states.
In a sense this should guarantee some scaling properties of quantum
entanglement, i.e. more-dimensional systems can be more
entangled.

\section{Ordering {\em vs.} local operations}
Entanglement measures enable us not only to decide whether a given state is entangled, but they  also allow us to conclude whether one state
is more entangled than another. In fact, any entanglement
measure can be used to induce an ordering on a set of quantum states.
However, it has been pointed out in Ref.~\cite{plenio} and analyzed by
many others \cite{miranowicz} that entanglement-induced
orderings for two different entanglement measures $E_1,E_2$
can differ. Even for the most commonly used measures of entanglement
\cite{miranowicz} there exists a pair of states $\omega_{AB}$ and
$\varrho_{AB}$ such that $E_1(\varrho_{AB})>E_1(\omega_{AB})$,
but $E_2(\varrho_{AB})<E_2(\omega_{AB})$.

In Ref.~\cite{ziman06} we addressed the question whether for a given
entanglement measure the ordering is preserved under the action of local operations
(without a classical communication). In a sense, we postulated an additional
axiom that should be fulfilled by a ``good'' entanglement measure.
There are several proposals for entanglement measures satisfying
the basic properties 1-4 from the above list. For example,
the entanglement of formation \cite{bvsw96}, the concurrence \cite{wootters98},
tangle \cite{ckw}, the relative entropy of entanglement \cite{vprk97}, the negativity
\cite{vw}, the squashed entanglement \cite{christandl04}, etc.
Certainly, the practical computability might be a non-trivial problem. In most cases
the optimization and the minimization can be accomplished only
numerically. For a two-qubit system the entanglement of formation
$E_f=\inf\sum_j p_j \tilde{S}_{vN}(\Psi_j)$,
the tangle $\tau=\inf\sum_j p_j \tilde{S}_L(\Psi_j)$
and the concurrence $C=\sqrt{\tau}$ are mutually closely related and they are straightforward to
to compute. We used the notation $\tilde{S}$ for the
corresponding entropy $S$ of the reduced state $\omega=\T_B\Psi$.
The infima are taken over all convex decompositions of the given state
$\varrho$ into pure states $\{\Psi_j\}$.  The indexes $vN$, $L$ stand for
the von Neumann entropy ($S_{vN}=-\T\varrho\log\varrho$) and
the linear entropy $S_L=2(1-\T\varrho^2)$, respectively. It was shown in
\cite{wootters98} that for two qubits $E_f=h(\frac{1}{2}[1+\sqrt{1-\tau}])$,
$\tau=C^2$ and $C=\max\{0,\sqrt{\lambda_1}-\sqrt{\lambda_2}
-\sqrt{\lambda_3}-\sqrt{\lambda_4}\}$, where $\lambda_j$ are decreasingly
ordered eigenvalues of the matrix
$R=\varrho(\sigma_y\otimes\sigma_y)\varrho^*(\sigma_y\otimes\sigma_y)$
and $h(x)=-x\log x-(1-x)\log(1-x)$ is the binary entropy.

\begin{figure}
\includegraphics[width=8cm]{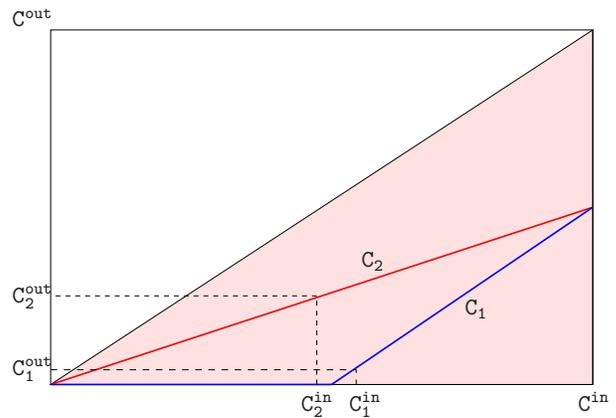}
\caption{The input/output diagram for the concurrence for
two families of states: 1) Werner states $\varrho_2=q\Psi_+ +(1-q)\frac{1}{4}I$
with $\Psi_+$ being a projector onto maximally entangled state
$|\psi_+\r=\frac{1}{\sqrt{2}}(|00\r+|11\r)$,
and 2) pure states $\varrho_1=|\psi\r\l\psi|$ with
$|\psi\r=\alpha|00\r+\beta|11\r$. We consider  the
depolarizing channel with $p=1/2$. The states from the
counterexample discussed in the paper are displayed and the
change in the ordering is visible. The region under the line
$C^{\rm out}=C^{\rm in}$ represents the allowed region that is
achievable by local channels. The concurrence is measured in
dimensionless units. \label{fig} }
\end{figure}

In our previous work \cite{ziman06} we have shown that
a stability of the entanglement-induced ordering is not compatible
with the listed axioms. A simple counter-example one can present involves
four qubits
divided into two groups.
Moreover, we have explicitly shown
that the ordering is not preserved for all
two-qubit measures providing one of the subsystems
is affected by the depolarizing
channel $\cE_p[\omega]=p\varrho+(1-p)\frac{1}{2}I$. The violation of the
ordering is depicted in the diagram on Fig.~1.  Based on
this explicit counter-example we can argue that there is no
(nontrivial) entanglement measure
$E$ that is stable under the action of local operations of the form $\cE\otimes\cI$,
where $\cE$ is a tracepreserving completely positive linear map on the
system $A$ only.

Let us consider the so-called unilocal channel of the form $\cE\otimes\cI$
and some entanglement measure $E$. The action of such local channel
can be expressed in the $[E_{\rm in},E_{\rm out}]$-diagram with
respect to a given measure of the entanglement $E$. Whenever we find that
for fixed values of $E_{\rm out}$ there exist more input values $E_{\rm in}$,
one can easily construct a suitable counter-example violating the condition of the
ordering-preservation
\be
E(\varrho_1)>E(\varrho_2)\, \Rightarrow\,
E(\varrho_1^\prime)\ge E(\varrho_2^\prime)
\ee
valid for all states $\varrho_1,\varrho_2$ and
$\varrho_j^\prime=\cE\otimes\cI[\varrho_j]$ ($j=1,2$).

More specifically. Let us define a ``horizontal fiber'' $\cF_h(E_{\rm out})$ to be
a set of all values of $E_{\rm in}$ such that there exists a state
$\varrho_{\rm in}$ with $E(\varrho_{\rm in})=E_{\rm in}$ and
$E(\cE\otimes\cI[\varrho_{\rm in}])=E_{\rm out}$. Whenever
$\cF_h(E_{\rm out})\cap\cF_h(E^\prime_{\rm out})\ne\emptyset$
for all pairs of possible values $E_{\rm out},E_{\rm out}^\prime$
and $\cF_h(E_{\rm out})\ne\cF_h(E_{\rm out})\cap\cF_h(E^\prime_{\rm out})
\ne\cF_h(E_{\rm out}^\prime)$,
the counter-example can be designed. Consider
$E_{\rm out}>E_{\rm out}^\prime$. Because of the nonempty intersection
of $\cF_h(E_{\rm out}),\cF_h(E^\prime_{\rm out})$, there
are states $\varrho_{\rm in}^{j}$ ($j=1,2$) with the same amount
of the initial entanglement
$E_{\rm in}^1=E_{\rm in}^2$, but different values of the final
entanglement $E_{\rm out}=E_{\rm out}^1>E_{\rm out}^2=E_{\rm out}^\prime$.
Moreover, it is possible to choose $\varrho^1_{\rm in}$
and $\varrho^2_{\rm in}$ to have different values of entanglement so that
the ordering is not preserved,
in particular, $E(\varrho_{\rm in}^1)<E(\varrho_{\rm in}^2)$.
Each unilocal channel $\cE$ determines a
set $S_\cE$ in the $[E_{\rm in},E_{\rm out}]$-diagram.
In particular, for $S_\cE$ forming some region (i.e. two-dimensional
geometrical object) the ordering is not preserved, because
there are values $E_{\rm out},E_{\rm out}^\prime$ for which
$\cF_h(E_{\rm out})\cap\cF_h(E^\prime_{\rm out})\ne\emptyset$.

The formal description presented in the above paragraph as well as
the particular analysis itself might be technically difficult.
In fact, the illustration of the set $S_\cE$ requires to evaluate
the entanglement for all possible states. However, intuitively
the situation expressed in Fig.~1 is not that complicated.

The observation that deserves special attention is that in order
to avoid the counter-examples of the above form the entanglement measure
and the transformation $\cE$ must have very specific (and very peculiar)
properties that are reflected in the $[E_{\rm in},E_{\rm out}]$-diagram.
If the possible values of $E_{\rm out}$ form a continuum (which is the case
for all the measures we use), then the corresponding
set $S_\cE$ must form a line. But this means, that either the
equally entangled states are always mapped into the equally entangled states,
or $S_\cE$ consists of horizontal and vertical lines. The corresponding
maps would be indeed interesting.

We started our discussion with the question whether
there exists an entanglement measure such that for {\em all} channels
$\cE\otimes\cI$ the induced ordering is preserved. However, the analysis
led us to another questions. Specifically,  for which channels a given entanglement
measure is preserved? Our conjecture is that essentially arbitrary local
channel affects the ordering. The only known exceptions are: 1) a unitary
channels ($E_{\rm out}=E_{\rm in}$), 2)
and the entanglement-breaking channels ($E_{\rm out}=0$).
Other ``entanglement-order-preserving'' channels would be
of interest {\em per se}. There is a strong evidence that such channels
do not exist. Consequently, it seems that the measures
stable under local operations should be discrete, i.e. the entanglement
can achieve only certain countable set of values. An example of
such measure is the trivial $\delta$-measure that answers the question
whether a given state is entangled, or not. Our statement holds modulo
this type of "discrete" entanglement measures.

\section{Maximal entanglement {\em vs.} local operations}
It is important to know how the entanglement behaves under the action of quantum dynamics
\cite{horodecki_zyckowski}.
For example, it is interesting to know whether local sources of decoherence
are relevant for a given quantum protocol based on entangled
states. In the previous section we have analyzed how the local operations
affect the entanglement-induced ordering. Positive answer to such question
would give us a strong tool how to analyze the effect of local noise
in general settings just by analyzing the behavior of the most entangled
states. Unfortunately, we have found that the situation is puzzling,
because it seems that essentially arbitrary unilocal channel
does not preserve the ordering whatever measure we choose. In this section
we will focus on a simpler question: How much can we learn from the analysis
of the dynamics of maximally entangled states?

In Ref.~\cite{ziman06} we concluded that maximally entangled
state remains most entangled also after the application of the
local transformation $\cE\otimes\cI$. Unfortunately, this statement is
not correct and there is a loophole in the proof
\cite{piani, ziman_piani}. Here is a simple counter-example.
Consider a system consisting of four qubits (the qubits $A,A^\prime$
on one side and $B,B^\prime$ on the other one) and a local map
$\cE_{AA^\prime}= \cP_0\otimes I + \cP_1\otimes\cA$,
where $\cP_j$ is defined as $\cP_j[X]=P_j X P_j$ ($P_j=|j\r\l j|$),
and $\cA[X]=\frac{1}{2}\T(X)I$. This transformation ``checks'' the state
of $A$ and either leaves $A^\prime$ unaffected, or it contracts its state
into a maximally mixed state. We will analyze the action
of such channel on two states:
1) $\varrho_1=\rho_{ABA^\prime B^\prime}=|0\r\l 0|_A \otimes |0\r\l 0|_B \otimes P^+_{A^\prime B^\prime}$, or
2) maximally entangled state
$\varrho_2=P^+=P^+_{ABA^\prime B^\prime} = P^+_{AB} \otimes P^+_{A^\prime B^\prime}$,
where $P^+_{AB}$ is a projector onto a maximally entangled state of qubits $A,B$,
and similarly for $P^+_{A^\prime B^\prime}$. The first of these states is
invariant under the action of $\cE_{AA\prime}\otimes\cI_{BB^\prime}$, i.e.
$\varrho_1^\prime=\varrho_1$, but
$\varrho^\prime_2=\cE_{AA^\prime}\otimes\cI_{BB^\prime}[P^+]=
\frac{1}{2}\varrho_1+\frac{1}{2} |1\r\l 1|\otimes|1\r\l 1|\otimes\frac{1}{4}I$.
The state $\varrho_2^\prime$ is, if entangled, for sure is strictly
less entangled than $\varrho_1^\prime$, i.e. ordering is not preserved
for an arbitrary measure of entanglement. The convexity guarantees that
$E(\varrho_2^\prime)\le \frac{1}{2}E(\varrho_1^\prime)<E(\varrho_1^\prime)$.

This result suggests that it is not straightforward to see how much
the analysis of dynamics of maximally entangled states can tell us about the
entanglement dynamics in general. On the other hand, in spite of the result
related to the entanglement-induced ordering, in the present case the maximality
is preserved for larger class of channels. Their
characterization is an open problem and will be analyzed elsewhere
\cite{ziman_piani}. An interesting feature that remains valid is that
all maximally entangled states are (under unilocal channels)
mapped into states with the same amount of entanglement \cite{ziman06}.
This holds for any measure of entanglement.

\section{Speculations and conclusions}
As a result of our analysis
we discovered new features and properties of entanglement measures. We found that
the ordering that implies statements such as ``one state is more/less entangled than another'' is not preserved
under the action of local operations. Moreover, such ordering is affected by all unilocal
operations except the unitary and the entanglement-breaking channels. Surprisingly
enough, we also found that the maximally entangled states might be
more fragile than ``less'' entangled states. This might sound
counter-intuitive, but in some realistic cases, in which the
systems are affected by a local noise, it could be better
to start with less (noise-dependent) entangled state in order to increase the success of the protocol.
Hence, the operational meaning of the property ``being more/less entangled'' is questionable.
Operationally, ``more entangled'' should be synonymous to ``having larger rate'' of success.
However, just a small modification of protocols (e.g. taking into account a local noise)
might change this interpretation. Hence,
does it make any sense to use the entanglement measures for the state ordering?
If not, then what are these measures good for?

Entanglement measures still provide us with very powerful tools
enabling us to decide the basic question, whether a given state
is entangled, or not. In fact, it is much simpler to compute
the concurrence of two qubits than to prove the (non)existence of a
separable decomposition. It might be that the idea of entanglement-induced
state ordering cannot be based on some entanglement measure. To introduce
such concept one should probably adopt different approach, in which
the stability with respect to local operations is fulfilled ``by the definition''.
Even in this case we have more options depending on the class
of operations we will consider. We can say that a state $\omega_1$ is more,
or equally entangled than a state $\omega_2$ ($\omega_1\succeq\omega_2$)
if and only if there exists a completely positive tracepreserving
linear operation $\cE_A\otimes\cE_B$ such that
$\omega_2=\cE_A\otimes\cE_B[\omega_1]$. This is compatible
with the fact, that entanglement can be only decreased by the action of  local
operations (LO). Alternatively, one can use the class of LOCC operations, or
stochastic LOCC (SLOCC) operations. Two states are equally entangled if
$\omega_1\succeq\omega_2$ and $\omega_2\succeq\omega_1$ simultaneously.
If two states are not equivalent, but $\omega_2\succeq\omega_1$, then
$\omega_2\succ\omega_1$. All these types of entanglement-based
orderings are, in principle, partial, i.e. not all states are comparable. For example, using
the SLOCC-ordering all two-qubit entangled states are equally entangled,
because they can be used for the teleportation. The LOCC-ordering is more
strict and for the LO-ordering pure states with different Schmidt
coefficients are not comparable. Intuitively, the most physical/operational
is the LOCC-based state ordering.

Recently, Kinoshita et al. in \cite{kinoshita}
analyzed compatibility of the LOCC-based ordering
under the action of local operations. They presented an example of
two states $\omega_1\succ\omega_2$ that are transformed
by a unilocal operation $\cE\otimes\cI$
(the so-called selective entanglement-breaking channels)
into $\omega_1^\prime,\omega_2^\prime$ such that
$\omega_{2}^\prime\succ\omega_1^\prime$. This explicit example
supports our conclusion about the existence of entanglement-induced
state orderings compatible with local operations,
because it shows that for an arbitrary entanglement measure
satisfying the the LOCC monotonicity
condition the entanglement-induced ordering is not preserved.
But, one can make even stronger conclusion that also the ``operational''
LOCC-based state ordering is not robust with respect to
local operations. It seems that there is no way how to
introduce a nontrivial entanglement-related state ordering
compatible with local operations. The only option is to use the
trivial $\delta$-measure, or some simple modification of it.

In the analysis of entanglement dynamics it is of interest to
specify times at which the entanglement disappears. Although
any particular dynamics depends on the initial state, these
``entanglement-breaking'' time instants $t_{sep}$ can be completely
characterized by the analysis of the maximally entangled state.
The channel is called entanglement-breaking $\cE$ if and only if
$\omega^\prime=\cE\otimes\cI[\omega]$ is separable for all initial
states $\omega$. It is sufficient to verify this property
for a maximally entangled state, i.e. whether $E(\cE\otimes\cI[P_+])=0$ \cite{03hsr}.
The local dynamics is given by a one-parametric set of completely positive
maps $\cE_t$. We have analyzed \cite{ziman_05} the general qubit
master equation generating semigroup dynamics. The qubit
semigroup  evolution is characterized by two time scales: the decoherence time
$T_{\rm decoherence}$ and the decay time $T_{\rm decay}$. What are the limits
on the entanglement decay? Which process is the fastest one? These questions
are not answered in \cite{ziman_05}, but all the necessary tools are derived
in that paper. It is known that in some cases $t_{sep}\to\infty$, but
what is the shortest possible decay time $t_{sep}$? The result is that there
is no limit and $t_{sep}$ can be arbitrarily small. For example, under
the action of a local depolarization process
$\cE_t[\varrho]=e^{-t/T}\varrho+(1-e^{-t/T})\frac{1}{2}I$ the maximally
entangled states evolves into the state
$\omega_t=e^{-t/T}P_+ +(1-e^{-t/T})\frac{1}{4}I$ (Werner states).
Hence, the entanglement vanishes for $t_{\rm sep}=T\ln 3$.
The parameter $T$ can be adjusted so that the
entanglement is destroyed in arbitrarily small time $t_{\rm sep}$.
In general, the vanishing decoherence rate guarantees the shortest
possible entanglement decay time, i.e. the process of entanglement decay
can be "infinitely" fast.

Let us get back to the status of entanglement measures. The main message of this contribution is that
the quantification of entanglement based
on entanglement measures define a state ordering that is not preserved
under the action of local operations. The interpretation of these measures should be reconsidered.
It seems that large values of entanglement measures characterize
the ``distance'' from the set of maximally entangled states, which is clearly
defined. Similarly, small values should correspond to states that are very far
(in the sense of entanglement) from the maximally entangled ones and
very close to the separable region of the state space.
The particular mathematical forms of these statements is not known,
but the meaning of entanglement degree could be hidden there.
The axiomatic entanglement measures can quantify different aspects
of quantum entanglement, or they can serve as bounds for particular
protocols. To understand the entanglement itself it is important to understand
the numbers we use to quantify this phenomenon. Thinking about the
relation between the state ordering, the entanglement measures, and the
robustness with respect to local operations, opens new
interesting conceptual questions deserving a deeper investigation.

\section*{Acknowledgements} We would like to thank to Marco Piani for
pointing out that there is a wrong statement in our earlier paper \cite{ziman06}. We thank him for  interesting
and encouraging discussions. This work was supported in part by the European
Union  projects QAP, CONQUEST, INTAS project number 04-77-7289, and  by the Slovak
Academy of Sciences via the project CE-PI, and by the projects
APVT-99-012304 and GA\v CR GA201/01/0413.


\end{document}